\documentstyle[12pt]{article}
\begin{document}
\title{Is the Universe Rotating?}
\author{B.G. Sidharth\\
International Institute for Applicable Mathematics \& Information Sciences\\
B.M. Birla Science Centre, Adarsh Nagar, Hyderabad - 500 063, India}
\date{}
\maketitle
\begin{abstract}
Numerous observations and studies suggest that the universe has some
sort of overall rotation. We consider this matter and provide a new
angle.
\end{abstract}
\section{Introduction}
Let us start by recapitulating some salient features of our solar
system. By and large, the sun and other planets of the solar system
are in the same plane. The planets revolve around the sun in the
same direction, which happens to be also the direction of rotation
of the sun. Moreover the plane of revolution of the planets is the
equatorial plane of the sun. The orbits of the planets are nearly
circular and are well spaced, that is with clear separation between
the planets. The same features are also observed of the satellites
of the planets. In the above we have not considered a few
exceptional cases, as they will not affect the following
considerations. Another interesting feature of the solar system is
that the angular momentum of all the planets put together is some
fifty times the angular momentum of the sun. This last observation
is apparently contrary to expectation. That is because the sun is so
much more massive than all the planets put together, and so, given
even its slow rotation, the expectation would be that its angular
momentum would be very high.\\
Based on these considerations, astronomers are broadly agreed on the
following scenario for the origin of the solar system \cite{hoyle}.
We start with a huge cloud, predominantly of Hydrogen which extends
beyond the present dimensions of the solar system. Such a cloud
would be spinning slowly and at the same time contracting under its
own weight. This would cause it to spin with increasing speed, till
eventually an equatorial disc of material is ejected. The same thing
would happen again and again. The planets condense out of such
discs. A mechanism that provides for the transfer of angular
momentum from the sun to the planets is found in magneto
hydrodynamic considerations. This model can also explain, broadly,
the composition of the various planets, for example the denser and
smaller planets which are closer, and the giant planets which are
farther and are made up of predominantly Hydrogen and Hydrogen
compounds. The details are omitted here as these are not relevant
for the sequel.
\section{"Spin"}
If we now step out of the solar system and take a look at what is
happening in the Milky Way galaxy, we find stars orbiting the
central nucleus of the galaxy, much like the orbiting planets. This
imparts an overall spin of the galaxy. It is also quite remarkable
that galaxies like the Milky Way have satellite galaxies which too
obey the broad types above like revolution in the same plane, like
the planetary orbits revolution in the same direction
and so on. All this undoubtedly points to a similar origin.\\
With this preface we consider the following. There appears to be a
broad self similarity across different scales in the universe. For
example the model of the atom, the solar system, the galaxy, as seen
above and even the satellite galaxies of galaxies. We will now argue
that it is as if there are different Planck constants at different
scales given by
\begin{equation}
h_1 \sim 10^{93}\label{8eaa1}
\end{equation}
for super clusters;
\begin{equation}
h_2 \sim 10^{74}\label{8eaa2}
\end{equation}
for \index{galaxies}galaxies and
\begin{equation}
h_3 \sim 10^{54}\label{8eaa3}
\end{equation}
for stars. And
\begin{equation}
h_4 \sim 10^{34}\label{8eaa4}
\end{equation}
for \index{Kuiper Belt objects}Kuiper Belt objects. In equations
(\ref{8eaa1}) - (\ref{8eaa4}), the $h_\imath$ play the role of the
\index{Planck}Planck constant, in a sense to be described below. The
origin of these equations is related to the following empirical
relations
\begin{equation}
R \approx l_1 \sqrt{N_1}\label{8eaa5}
\end{equation}
\begin{equation}
R \approx l_2 \sqrt{N_2}\label{8eaa6}
\end{equation}
\begin{equation}
l_2 \approx l_3 \sqrt{N_3}\label{8eaa7}
\end{equation}
\begin{equation}
R \sim l \sqrt{N}\label{8eaa8}
\end{equation}
and a similar relation for the KBO (\index{Kuiper Belt
objects}Kuiper Belt objects)
\begin{equation}
L \sim l_4 \sqrt{N_4}\label{8eaa9}
\end{equation}
where $N_1 \sim 10^6$ is the number of superclusters in the
\index{Universe}Universe, $l_1 \sim 10^{25}cms$ is a typical
supercluster size, $N_2 \sim 10^{11}$ is the number of
\index{galaxies}galaxies in the \index{Universe}Universe and $l_2
\sim 10^{23}cms$ is the typical size of a galaxy, $l_3 \sim 1$ light
years is a typical distance between stars and $N_3 \sim 10^{11}$ is
the number of stars in a galaxy, $R$ being the radius of the
\index{Universe}Universe $\sim 10^{28}cms, N \sim 10^{80}$ is the
number of \index{elementary particles}elementary particles in the
\index{Universe}Universe and $l$ is a typical elementary particle
\index{Compton wavelength}Compton wavelength and $N_4 \sim 10^{10},
l_4 \sim 10^5 cm$, is the dimension of a typical KBO (with
\index{mass}mass $10^{19}gm$ and $L$ the width of the Kuiper Belt
$\sim 10^{10}cm$
cf.ref.\cite{cu}).\\
The size of the \index{Universe}Universe, the size of a supercluster
etc. from equations like (\ref{8eaa5})-(\ref{8eaa9}), as described
in the references turn up as the analogues of the \index{Compton
wavelength}Compton wavelength. For example we have
\begin{equation}
R = \frac{h_1}{Mc}\label{8eaa10}
\end{equation}
where $M$ is the mass of the universe. One can see that equations
(\ref{8eaa1}) to (\ref{8eaa10}) are a consequence of
\index{gravitation}gravitational orbits (or the Virial Theorem) and
the conservation of angular momentum viz.,
\begin{equation}
\frac{GM}{L} \sim v^2, M v L = H\label{8eaa11}
\end{equation}
(Cf.refs.\cite{BGSSU1,BGSSU2}), where $L,M,v$ represent typical
length (or dispersion in length), \index{mass}mass and velocities at
that scale and $H$ denotes the scaled \index{Planck}Planck constant.
As another example, if we use the figures for the mass $(\sim
10^{44}gm)$, velocity $v (\sim 300 km/sec)$ and radius $L$ of a
galaxy $(\sim 10^{24})$, we can recover (\ref{8eaa2}).\\
We can indeed give a rationale for (\ref{8eaa1}) from a slightly
different point of view by considering in the equation for the spin
in the linearized General Relativistic case, the universe itself
with $N \sim 10^{80}$ particles.\\
In the case of the electron, it was shown \cite{cu} that the spin
was given by,
\begin{equation}
S_K = \int \epsilon_{klm} x^lT^{m0} d^3 x = \frac{h}{2}\label{eb1}
\end{equation}
where the domain of integration was a sphere of radius given by the
Compton wavelength. If this is carried over to the case of the
universe, we get from (\ref{eb1})
\begin{equation}
S_U = N^{3/2} h \approx h_1\label{eb2}
\end{equation}
where $h_1$ which is the same as in (\ref{8eaa10}), and $S_U$
denotes the counterpart of electron spin (Cf.ref.\cite{cu}). In
deducing (\ref{eb2}), use has been made of (\ref{8eaa8}).\\
With $N \sim 10^{80}$, the number of elementary particles in the
universe $h_1$ in (\ref{eb2}) turns out to be the spin of the
universe itself in broad agreement with Godel's spin value for
Einstein's equations \cite{godel,carneiro}. Incidentally this is
also in agreement with the Kerr limit of the spin of the rotating
Black Hole. Further as pointed out by Kogut and others, the angular
momentum of the universe given in (\ref{eb2}) is compatible with a
rotation from the cosmic background radiation anisotropy
\cite{carneiro}. Finally it is also close to the observed rotation
as deduced from anisotropy of cosmic electromagnetic
radiation as reported by Nodland and Ralston and others \cite{ralston,kogut}.\\
In the above $h_1 \sim 10^{93}$ and we immediately have
\begin{equation}
R = \frac{h_1}{Mc}\label{eb3}
\end{equation}
whiis (\ref{8eaa10}) and where $R$ the radius of the universe is the
analogue of the particle Compton wavelength in the macro context and
$M$ is the mass of the universe. This itself substantiates our claim
that the entire universe with its constituents rotates in the above
self similar
sense.\\
There is another way of looking at this. Let us consider a Kerr
Black Hole. As is known its horizon is given by \cite{ruffinizang}
\begin{equation}
r = \frac{GM}{c^2} + (\frac{G^2M^2}{c^4} - a^2)^{1/2}\label{15}
\end{equation}
where $a$ is the angular momentum of the rotating Black Hole and is
given by
\begin{equation}
a = MRc\label{16}
\end{equation}
where $R$ is the radius of the Black Hole. We can see from
(\ref{15}) and (\ref{16}) that the radius $R$ is given by
\begin{equation}
R \sim \frac{GM}{c^2}\label{17}
\end{equation}
Indeed it is well known that (\ref{17}) holds good for the universe
as a whole \cite{tduniv,weinberggravcos}. In other words we can
indeed treat the universe as a Kerr Black Hole exactly as in the
discussion leading to (\ref{eb3}).
\section{Remarks}
We would expect that such a rotation of the universe would lead to a
anisotropy in the cosmic microwave background, and as pointed out
above, this indeed seems to be the case. Additionally Palle
\cite{palle} too argues from WMAP data for such an anisotropy.\\
We could also expect some magnetic effects due to this rotation.
This would follow from a relation first put forward by Blackett
\cite{blackett} viz.,
\begin{equation}
\mu \sim \left(G^{\frac{1}{2}} / c\right) L\label{AA}
\end{equation}
In (\ref{AA}) $\mu$ is the magnetic movement while $L$ is the
angular momentum of a cosmic object. This has been discussed in
detail by De Sabbata \cite{sabbata}. We note that (\ref{AA}) gives a
huge contribution for the magnetism if we use the value $h_1$ from
(\ref{eb3}) for $L$. However as can be calculated it leads to an
average magnetism density $\sim 10^{-5} gauss$. Feeding into
(\ref{AA}) the values for a typical galaxy, we get instead a field
density $\sim 10^{-12} gauss$. It may be mentioned that such cosmic
(including galactic magnetism) has been detected
\cite{parker,kul}.\\
Finally it may be mentioned that many years ago in the context of
Mach's Principle, Pietronero too investigated this problem
\cite{nero}.
\newpage
\noindent {\bf {\large APPENDIX}}\\ \\
We have referred to the Blackett formula and
the work of Sabbata and others above. Let us look at this briefly.\\
L. Pietronero many years ago argued that an interpretation of
rotation from the Mach point of view implies the following:
\begin{equation}
\Omega (r) = \Omega_0 / \sqrt{1 + \Omega^2_0 r^2 / c^2}\label{a1}
\end{equation}
This means that for $r \to \infty$ we have
\begin{equation}
\Omega (r) r \to c\label{b1}
\end{equation}
In the above $\Omega$ is given by
\begin{equation}
\Omega = \frac{Mc^2}{H} (H = Mcr)\label{c1}
\end{equation}
This indeed is in agreement with our "scaled" Planck constant
approach with $H$ playing the role of $h$ and $\Omega$ playing the
role of the frequency of microphysics.\\
We have also alluded to Blackett's formula. This was conjectured by
Blackett when he noticed that the observed ratio of the magnetic
dipole movement over the angular momentum was nearly the same for
the Earth, the Sun, the star 78 Virginis. It was also found that
this ratio holds good for other stars as well including White Dwarfs
and Pulsars. Indeed it is interesting that this formula too has an
analogue in microphysics -- with the gyromagnetic ratio of the
electron. Indeed Sabbata too looked at this problem from the point
of view of the Dirac Large Number hypotheses which leads to a
gravitational constant varying inversely with time.\\
Finally it may be mentioned that Palle, whose work has been referred
to above has analyzed the WMAP data, particularly its aspects which
show an asymmetry and anisotropy in cosmic microwave background
fluctuations. He has argued that the covariant and gauge invariant
treatment of density fluctuations formulated by authors like Ellis
can explain the asymmetric and anisotropic WMAP data by including
the rotation of the universe. He claims that the spatial gradients
of the density are proportional to the spin of the universe.

\end{document}